# GUIDANCE, NAVIGATION AND CONTROL OF ASTEROID MOBILE IMAGER AND GEOLOGIC OBSERVER (AMIGO)

Greg Wilburn,[*] Himangshu Kalita,[†] Aman Chandra,[†] Stephen Schwartz,[‡] Erik Asphaug,[±] and Jekan Thanga[¥]

The science and origins of asteroids is deemed high priority in the Planetary Science Decadal Survey. Major scientific goals for the study of planetesimals are to decipher geological processes in SSSBs not determinable from investigation via in situ experimentation, and to understand how planetesimals contribute to the formation of planets. Ground based observations are not sufficient to examine SSSBs, as they are only able to measure what is on the surface of the body; however, in situ analysis allows for further, close up investigation as to the surface characteristics and the inner composure of the body. To this end, the Asteroid Mobile Imager and Geologic Observer (AMIGO) an autonomous semi-inflatable robot will operate in a swarm to efficiently characterize the surface of an asteroid. The stowed package is 10×10×10 cm (equivalent to a 1U CubeSat) that deploys an inflatable sphere of ~1m in diameter. Three mobility modes are identified and designed: ballistic hopping, rotation during hops, and up-righting maneuvers. Ballistic hops provide the AMIGO robot the ability to explore a larger portion of the asteroid's surface to sample a larger area than a stationary lander. Rotation during the hop entails attitude control of the robot, utilizing propulsion and reaction wheel actuation. In the event of the robot tipping or not landing upright, a combination of thrusters and reaction wheels will correct the robot's attitude. The AMIGO propulsion system utilizes sublimate-based micro-electromechanical systems (MEMS) technology as a means of lightweight, low-thrust ballistic hopping and coarse attitude control. Each deployed AMIGO will hop across the surface of the asteroid multiple times. Individual actuation of each microvalve on the MEMS chip provides control torque for rough attitude control with only slight alteration to the hop path en-route to its destination. For optimal use of instrumentation, namely the top mounted stereo cameras utilized in local surface mapping and navigation planning, the robot must remain as upright as possible during data acquisition. Should AMIGO land in an improper orientation, thrusters and reaction wheels will attempt to correct the positioning. Several inflatable structures will be evaluated including a soft inflatable and an inflatable that rigidizes under UV light. The inflatable will be compared under operational scenarios to determine if it produces disturbances torque and an un-steady view for the stereo cameras. Future work is focused on raising the TRL by real world testing system performance and utilizing hardware-in-the-loop simulation models. The thruster assembly can be evaluated on a test stand mounted inside a vacuum chamber. To simulate milli-gravity, the entire robot will be analyzed in either parabolic flight tests or in buoyancy chambers. A combination of experimentation will validate simulations and provide insight in areas to improve on the design and control algorithms for milli-gravity asteroid surface environments.

---

[*] Masters Student, Space and Terrestrial Robotic Exploration Laboratory, University of Arizona, Tucson, Arizona 85721.
[†] PhD Student, Space and Terrestrial Robotic Exploration Laboratory, University of Arizona, Tucson, Arizona 85721.
[‡] Postdoctoral Research Associate, Lunar and Planetary Laboratory, University of Arizona, Tucson, Arizona 85721.
[±] Professor, Lunar and Planetary Laboratory, University of Arizona, Tucson, AZ, 85721
[¥] Assistant Professor, Space and Terrestrial Robotic Exploration Laboratory, University of Arizona, Tucson, Arizona 85721.

# INTRODUCTION

The 2013 Planetary Science Decadal Survey highlights a few questions small solar system bodies (SSSBs), particularly asteroids, can answer: What are primordial sources of organic matter? What are the initial stages and processes of solar system formation? What makes up the matter that coalesced into larger bodies [1]? Even within the similar size and mass, asteroids can be quite diverse and are thought to hold knowledge on the formation and evolution of the solar system. The majority of the known asteroids are C-type for their mostly carbonaceous makeup, S-type with iron and magnesium rich silicates, and M-type with metallic rich surfaces [2]. Earth based telescopes can provide much information on the general composition and surface of an asteroid, but in-situ analysis is required to obtain details on the inner structure and other complex geological inquiries.

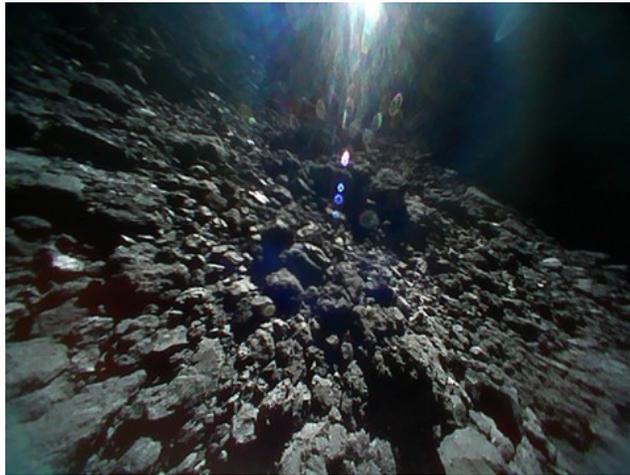

**Figure 1: Surface of Ryugu from MINERVA-II 1B (Credit: JAXA, University of Tokyo et al.).**

Some of the basic unknowns of asteroids are the cohesion of the outer surface regolith (fine dust particles), electrostatic forces [3], thermal effects, and geologic structure. The internal structure is thought to be a range of "fluffy aggregates" for smaller asteroids to "rubble piles" to fully differentiated, solid rock. Future, large scale surface landers need more information on the structure of asteroids to safely land, conduct science experiments and perform In-Situ Resource Utilization (ISRU).

Small robots that can sample the asteroid's surface at multiple locations provides more robust science data. The micro- to milli-gravity of asteroids provides important challenges to achieving mobility. For asteroid 101995 Bennu sized object, with equatorial radius ~262 meters and mass ~$7 \times 10^{10}$ kg, local gravitational attraction is 10 micro-g with an escape velocity of 0.19 m/s. A 1U CubeSat, 10×10×10 cm in volume and 1.0 kg in mass, is sufficient to perform significant science experiments, such as seismic sensing, micro-imaging, and electric field measurements.

# RELATED WORK

There are a few methods to obtain mobility across an asteroid's surface: wheeled rovers, internally actuated hops, mechanical devices, and thrusters. Traditional rovers are well suited to large bodies, such as the Moon and Mars, where local gravity is comparable to the Earth's. Wheels are able to turn and overcome small bumps with gravity holding the robot to the surface; however, in milli- and micro-gravity environments, the rotation of a wheel will impart a large enough delta-v to lift the robot off the surface and potentially into an escape trajectory [4].

Internally actuated devices typically rely on spinning up and braking a reaction wheel. A benefit to this system is the actuators are shielded from the surface regolith, extending their lifetime and limiting the probability of failure. The dynamics of the surface regolith must be well understood for accurate prediction of hopping dynamics, as the force transferred to the robot is dependent on robot-regolith interaction. Hedgehog is one such development by NASA JPL and Stanford, with three flywheels and external spikes to tumble for short distances and hop for more distant targets [5, 6]. Another is the Gyrover that contains spinning flywheels attached to a two-link manipulator [31]. A recent successful example of this concept is JAXA's MINERVA-II 1A and 1B landers, as they landed and hopped around the surface of Ryugu and transmitted images (Figure 1).

One type of mechanical hopping is by the use of a spring mechanism, a direct reactive force pushing the robot from the surface. The Canadian Space Agency developed the Micro-hopper for traversing Martin terrain, though with a limitation of only one hop per day due to the time to reform the shape memory alloy [7]. Another technique for hopping developed by Plante and Dubowsky at MIT utilize Polymer Actuator Membranes (PAM) to load a spring. The system is only 18 grams and can enable hopping of Microbots with a mass of 100 grams up to a 1 m [32].

Another example is SPIKE, a 75 kg spacecraft-hopper that embeds science instruments into regolith via a boom connected to the robot in free fall [8]. Vibrating the boom causes cohesion with regolith to be broken and the spacecraft is free to hop to another location. Again, mechanical hoppers have a reliance on surface characteristics, which are not well constrained and vary asteroid to asteroid.

Thrusters allow for mobility independent of surface characteristics, though exhaust may cause interference with the electrically charged, organic regolith and kick up dust in the process. Another example is the Sphere-X, a spherical robot that hops using chemical propulsion and is intended for exploring in higher gravity of 1.0 m/s$^2$ and higher [9-12]. This system, however, relies on reaction wheels to provide attitude control, as the thruster is used only for launching the robot. A thruster with multiple nozzles is required for pointing authority for smaller robots with less volume and mass for angular momentum transfer devices.

**AMIGO MISSION CONCEPT**

The Asteroid Mobile Imager and Geologic Observer (AMIGO) (Figures 2-3) would be deployed at multiple locations around the surfaces of small bodies and provide stereo imaging from vantage points ~1 m above the surface, close-up geologic imaging, and seismic sensing. Each lander contains a 1 m diameter inflatable for communicating at 256 Kbps and tracking and an on-board propulsion system to perform surface hopping. The inflatable is a critical multi-functional element of AMIGO as it addresses issues of tracking a small lander on the surface of an asteroid using an overhead mothership and the needs for high-communication bandwidth to transmit surface videos and images. The concept of operations is shown in Figure 4.

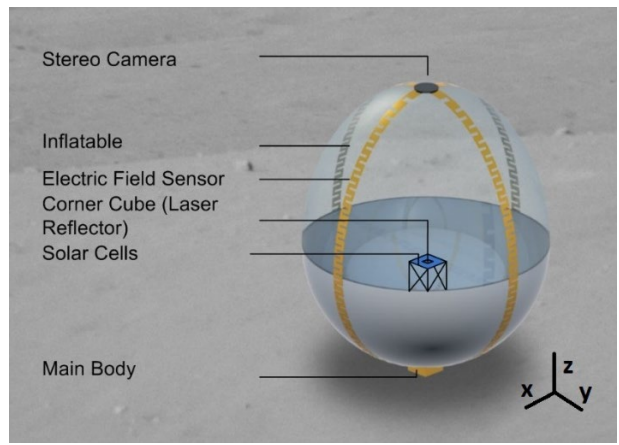

**Figure 2: AMIGO Overview**

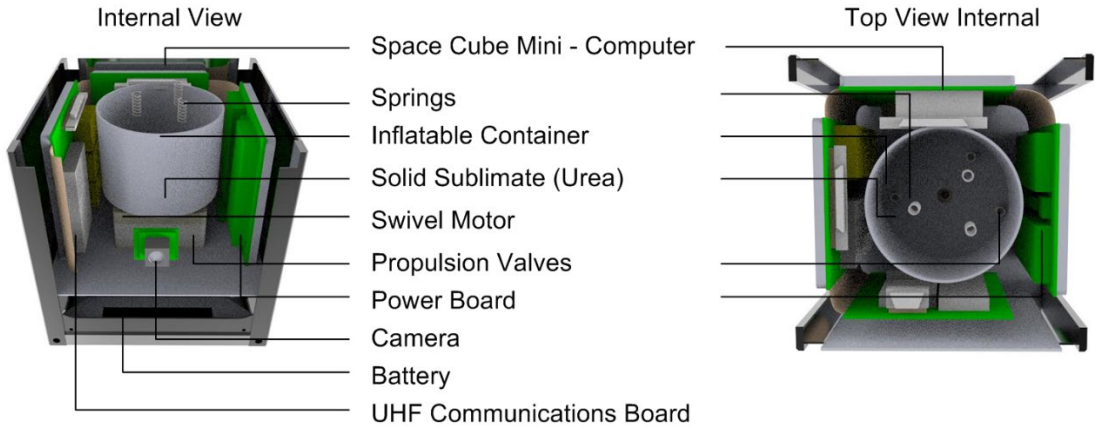

**Figure 3: AMIGO Internals**

Each AMIGO is deployed from a mother spacecraft (Figure 4). During descent, the robot inflates from its stowed 1U state. Upon landing, initial context is determined for where the robot is on the surface. This is done by both on board imaging and tracking from the mother. The inflatable portion provides a tracking target, as smaller robots may not be large enough to be tracked. From there, the science mission is conducted. For the AMIGO lander, there are five science goals:

1. Determine local surface hardness and compliance
2. Acquire seismic data constraining the geologic competence of the asteroid
3. Acquire micro-imaging of fine geologic structure from diverse locations
4. Detect images of thermal fatigue of surface rocks
5. Measure electric fields and properties of surface regolith

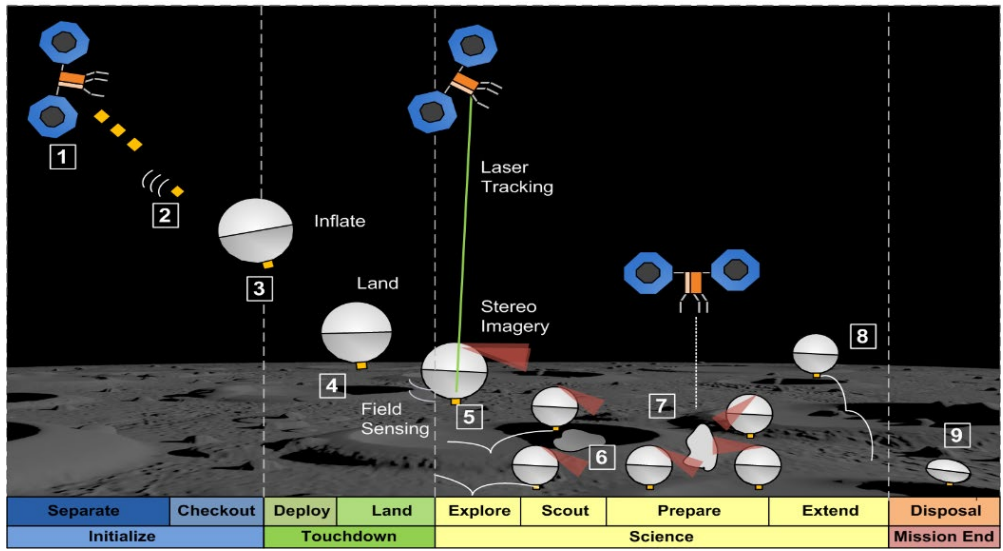

**Figure 4: AMIGO Concept of Operations.**

Each of these science goals seeks to fill a current knowledge gap in the characteristics of asteroids. For example, the proposed NASA Asteroid Redirect Mission was to retrieve a boulder from the surface of a near Earth asteroid and return the sample for further analysis [13]. Currently, the dynamics of how to extract a boulder from the surface of an asteroid is an open problem. The issue is as fundamental as Newton's Third Law; if one aims to pull a three-ton boulder from the asteroid surface, the spacecraft must exert three tons on the asteroid. Will the asteroid and boulder have enough cohesive strength to not completely fall apart? Seismic sensors and close-up geologic sensors will provide this information. The top-mounted camera provides context to determine local areas of interest and potential locations to traverse to (Figures 5-6).

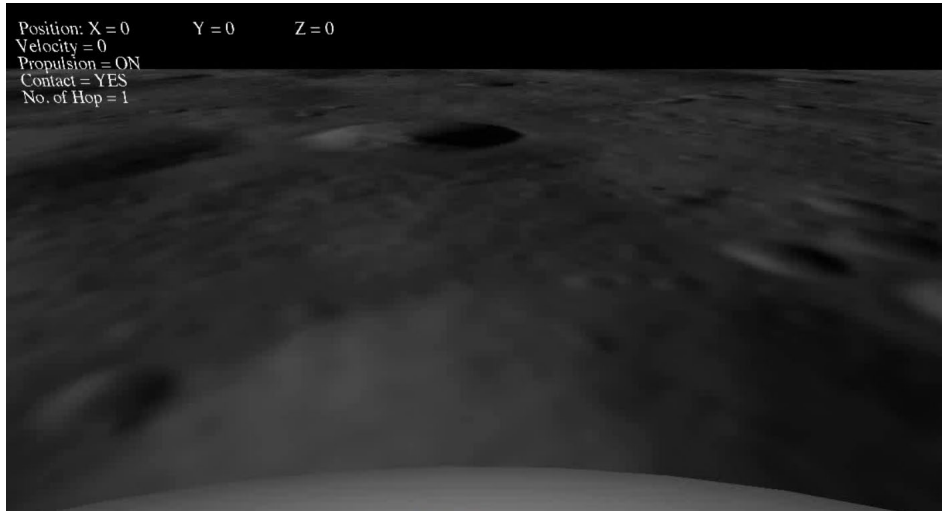

**Figure 5: AMIGO View from Asteroid Surface**

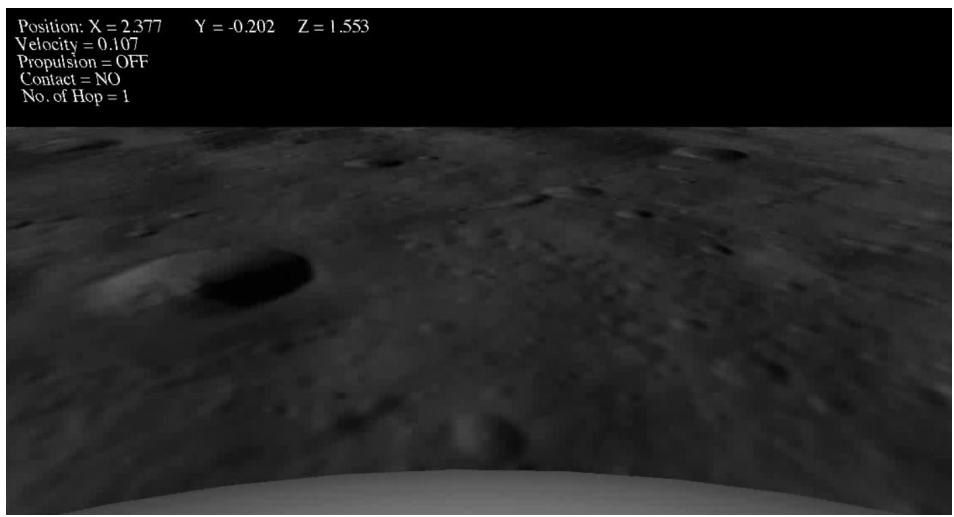

**Figure 6: AMIGO View from Top of Hop**

The characterization of surface regolith of asteroids is vital to the success of future lander missions and the further understanding of the composition of asteroids. For instance, it is theorized that planetesimals often impacted with each other and either obliterated into fine dust and small clumps or aggregated together. In either case, fine grains are created. For intact planetesimals, this dust accreted to the surface and became the surface regolith. However, that regolith may not have the same compositions as the asteroid itself due to being a combination of multiple meteoric impact events. In situ analysis will aid in the understanding of the surface of asteroids in this regard. A large reason for the concept of AMIGO is to add to the current base of knowledge for the surface characteristics of asteroids for use in future lander missions. The familiarity with asteroid surfaces gained by lower cost missions will lay the foundation for, say, a Discovery class mission to be more successful due to limiting the unknowns in the geology dynamics of asteroids.

**CONTROL ACTUATORS**

Motion of the robot is obtained by two types of actuators: an array of micro-thrusters and a reaction wheel. The thruster array is a MEMS chip of micro-nozzles based on sublimate cold gas propulsion (Figure 7) [14]. The purpose of the propulsion system is to provide thrust to lift off the surface of the asteroid and perform a hop to a new location, and to control the robot's attitude during the hop to ensure safe, upright landing. There are 8 nozzles in total, each capable of delivering 30 micro-Newtons of thrust. The thruster chip provides control torque on the x and y body axes, defined in Figure 2. The geometry of the nozzles allows for three modes of actuation to control one axis rotation: actuating the inner nozzle, outer nozzle, or both nozzles at the same time.

As the propulsion system is a chip with thrust only out of plane due to micro-fabrication limits, a reaction wheel is needed to control the z body axis of the robot.

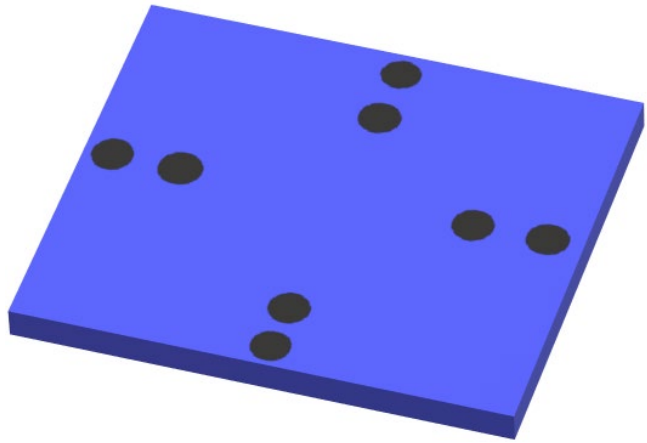

**Figure 7: Thruster Chip**

**Thrusters**

Propellant is stored as a sublimate in a heat-controlled storage chamber (Figure 8). The sublimate vapor pressure of the propellant is the chamber pressure, analogous to other cold gas and liquid evaporation systems [15-20]. A main valve that provide the main sealing pressure opens to allow for flow to downstream nozzles. Each nozzle is actuated by a simple thruster valve.

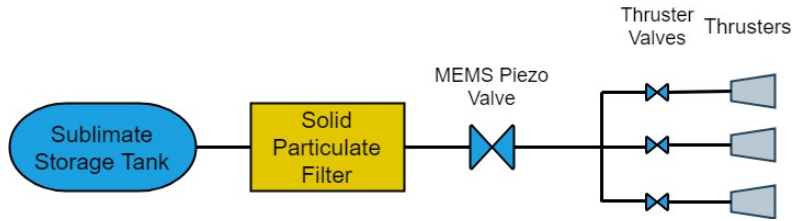

**Figure 8: Propulsion System Block Diagram**

**Reaction Wheels**

A small reaction wheel is required for z axis control. A commercial off the shelf solution from MAI is taken as a representative solution. The reaction wheel is 8.1 mm in radius, $2.25 \times 10^{-5}$ kg-m$^2$ inertia, and max spin of 5,000 rpm. This provides sufficient control authority for the robot. By changing the spin rate of the reaction wheel, the change in angular momentum is transferred to the robot to conserve angular momentum.

**BALLISTIC HOPPING**

The dynamic equations of motion of the robot in the asteroid's body fixed coordinate system is expressed as:

$$\ddot{r} + 2\omega \times \dot{r} + \omega \times (\omega \times r) + \dot{\omega} \times r = g + d + u \qquad (1)$$

where, $r$ is the position vector, $\dot{r}$ and $\ddot{r}$ are the first and second derivative of the position vector, $\omega$ is the angular velocity vector of the asteroid, $g$ is the gravitational acceleration, $d$ is the disturbance acceleration such as SRP and third body perturbations, and $u$ is the control acceleration. Also, the asteroid is considered to have a fixed angular velocity, so $\dot{\omega}$ is equal to zero. Although gravity in smaller bodies are weaker than on Earth, it still is the dominant force on robots. The polyhedral model is the most accurate gravity model for smaller irregular bodies which leverages the divergence theorem to exactly model the gravitational potential ($U$), gravitational acceleration ($g = \nabla U$), gradient ($\nabla \nabla U$) and Laplacian ($\nabla^2 U$) of a constant density polyhedron as a summation over all facets and edges of the surface mesh.

The rover needs to hop from rest at position $r_0$ with velocity $v_0$ and impact at position $r_f$ with velocity $v_f$. The problem of computing the launch velocity, $v_0$ to intercept a target location, $r_f$ at time $\tau = t_f - t_0$ is the well-known "Lambert orbital boundary-value problem" and efficient numerical solutions for different types of gravity fields are available. For the case of asteroids with irregular gravity field, a simple shooting method is used to calculate the launch velocity to successfully impact a target location [21].

Figure 9 shows the trajectory of the robot from its initial position to the final position on asteroid Itokawa. This asteroid was used due to the similar size and mass to Bennu with a high-fidelity shape model for simulation purposes. Two waypoints were added in between. The robot is able to find the initial hopping velocities required to reach its target location from its initial location successfully visiting the waypoints in between. Close targets are able to efficiently be reached by just one hop, but longer excursions require less efficient multi-hop schemes.

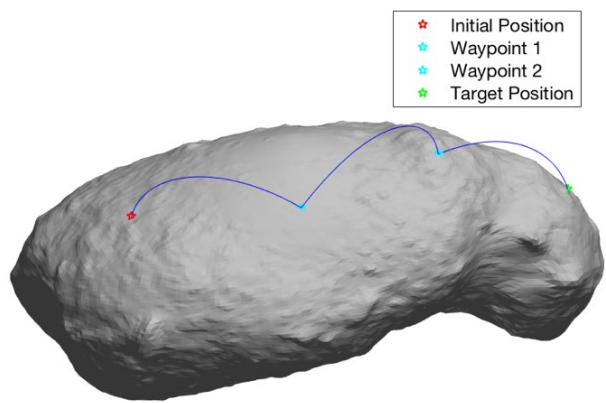

**Figure 9: Ballistic Hopping Trajectories from an Initial Position to a Final Position with Two Waypoints In-between on Itokawa**

## ATTITUDE CONTROL DURING HOPS

Quaternions are used as the attitude for the hopper. Quaternions are defined as

$$q = [q_1 \ q_2 \ q_3 \ q_4]^T \tag{2}$$
$$q = [q_v \ q_4]^T$$

Considering the equations of motion of a rigid spacecraft,

$$\dot{q}_v = \frac{1}{2}[\tilde{q}_v + I_{3x3}q_4]\omega \tag{3}$$

$$\dot{q}_4 = -\frac{1}{2}q_v^T\omega \tag{5}$$

$$J\dot{\omega} = -\tilde{\omega}J\omega + sat(u) + L \tag{6}$$

Where $\omega \in R^3$ is the angular velocity vector, $I_{3x3}$ is the 3×3 identity matrix, the ~ is the skew symmetric operator, $J \in R^{3x3}$ is the spacecraft inertia matrix, $u \in R^3$ is the control torque, the sat function corresponds to the saturation control modes, and $L \in R^3$ is the disturbance torque. Quaternions maintain structure through one constraint equation,

$$q_1^2 + q_2^2 + q_3^2 + q_4^2 = 1 \tag{7}$$

Meaning quaternions lie on a three-dimensional hypersphere in four-dimensional space, $q \in S^3$. It is assumed that the attitude q and angular velocity $\omega$ are known.

It is desired to design a control system to drive the spacecraft to the identity attitude ($q_v = 0$) with no angular velocity ($\omega = [0\ 0\ 0]^T$). To accomplish this, an adaptive sliding mode controller (ASMC) is used for its properties of robustness and disturbance rejection [22].

### ASMC Design

The saturated control authority is based on limitations from the thrusters; i.e. the discretized nozzles are only capable of providing certain amounts of torque on the robot. Considering the three saturation modes of minimum, medium, and maximum control torque, the saturation function is defined as

$$sat(u) = \begin{cases} 0, & abs(u_i) < u_{min} \\ u_{min}, & u_{min} \leq u_i < u_{mid} \\ u_{mid}, & u_{mid} \leq u_i < u_{max} \\ u_{max}, & u_{max} \leq u_i \end{cases} \tag{8}$$

Where $u_i$ are the required control torque components in the robot body axes and $abs(u_i)$ is the absolute value of the torque component. This saturation case is mirrored for negative torque, meaning rotation in the opposite direction. For the z-axis case, there is simply a maximum saturation, as the reaction wheel has continuous spin rates from no spin to its rated maximum. The input control torque must always satisfy these saturation conditions.

Drawing from the results from Zhu et. al [23], the control law is defined as

$$u = -\tau S - \sigma sgn(S) - u_a \tag{9}$$

Where $\tau = diag(\tau_i), \tau_i > 0$ and $\sigma = diag(\sigma_i), \sigma_i > 0$ are design parameters to reduce chattering upon reaching the sliding surface, S is the sliding surface ($S \in R^3$)

$$S = \omega + k_1 q_v \tag{10}$$

With design scalar $k_1 > 0$, $sgn(S)$ is the sign function

$$sgn(S_i) = \begin{cases} -1, S_i < 0 \\ 1, S_i \geq 0 \end{cases} \tag{11}$$

And adaptive controller $u_a$

$$u_a = \beta k_2 k_3 \|\xi\| \frac{\|S\|}{S} \tag{12}$$

$$\xi = [q_v \ \omega \ 1] \tag{13}$$

The update control laws are functions of time,

$$\dot{k}_2 = \beta k_2^3 k_3 \|\xi\| \|S\|, k_2(0) > 0 \tag{14}$$

$$\dot{k}_3 = p\|\xi\| \|S\|, k_3(0) > 0 \tag{15}$$

With β > 1 and p > 0 are design parameters. The adaptive portion of the control law rejects inertia uncertainties and external disturbances. From [23], it is shown that the control law Equation (9) is asymptotically stable for some bounded model uncertainties and disturbance torques. In actuality, due to the discontinuous nature of the control actuator, the controllable subspace may not be driven to S = 0. Rather, the robot will likely chatter around the desired state. The use of the "boundary layer" approach in [23] is neglected due to the lack of fine pointing control offered by the thrusters; no chattering can be reduced due to the minimum impulse bit.

**Disturbances: SRP and Gravity Field**

The most important perturbing forces for the robotic hopper are solar radiation pressure (SRP) and gravity effects, namely oblateness and ellipticity [24]. Formulating the solar radiation pressure in terms of a potential function,

$$R_{SRP} = \frac{-\beta}{d_{ast}^2} \hat{d}_{ast} \cdot \hat{r} \tag{16}$$

Where $\hat{d}_{ast}$ is the vector from the sun to the asteroid, $\hat{r}$ is the vector of the orbiter from the asteroid center of mass in the asteroid reference frame, and β is quantified as

$$\beta = (1 + \rho) \cdot \left(\frac{A}{m}\right) P_\varphi \tag{17}$$

Where ρ is the albedo of the spacecraft, A is the cross-sectional area, m is the mass, and $P_\varphi$ is the solar radiation constant. The asteroid gravitational effects are captured in the oblateness $C_{20}$ and ellipticity $C_{22}$ coefficients in a 2nd degree gravity perturbation,

$$R_m = \frac{-\mu}{2r^3} C_{20}[1 - 3(\hat{r} \cdot \hat{p})^2] + \frac{3\mu}{r^3} C_{22}[(\hat{r} \cdot \hat{s})^2 - (\hat{r} \cdot \hat{q})^2] \tag{18}$$

Where μ is the gravitational parameter of the asteroid, $\hat{p}$ is the maximum principal inertia axis, $\hat{q}$ is the intermediate principal inertia axis, and $\hat{s}$ is the minor principal inertia axis of the asteroid. Higher order effects are ignored. In an inertially fixed reference frame with the origin at the center of mass of the asteroid, the equations of motion of a satellite in orbit about an asteroid can be defined as

$$\ddot{\vec{r}} = \frac{\partial U}{\partial \vec{r}} + \frac{\partial R_{SRP}}{\partial \vec{r}} + \frac{\partial R_m}{\partial \vec{r}} \tag{19}$$

Where the potential U is

$$U = \frac{\mu}{r} \tag{20}$$

The disturbing accelerations modify the Keplerian motion, as seen in Equation (19). To compute the non-averaged disturbing accelerations,

$$\vec{a}_d = \vec{a}_{d_{SRP}} + \vec{a}_{d_m} = \frac{\partial R_{SRP}}{\partial \vec{r}} + \frac{\partial R_m}{\partial \vec{r}} \qquad (21)$$

Considering the SRP perturbations, the disturbing acceleration calculation is simple,

$$\vec{a}_{d_{SRP}} = \frac{-\beta}{d_{ast}^3}\hat{d}_{ast} \qquad (22)$$

To formulate the mass distribution acceleration, the disturbance is split into the oblateness and ellipticity effects for easier derivation.

$$\vec{a}_{d_{20}} = \frac{\partial}{\partial \vec{r}} \frac{-\mu}{2r^3} C_{20}[1 - 3(\hat{r} \cdot \hat{p})^2] \qquad (23)$$

$$\vec{a}_{d_{20}} = \frac{-\mu}{2} C_{20}[1 - 3(\hat{r} \cdot \hat{p})^2] \frac{\partial}{\partial \vec{r}} \frac{1}{r^3} + \frac{3\mu}{2r^3} C_{20} \frac{\partial}{\partial \vec{r}} (\hat{r} \cdot \hat{p})^2 \qquad (24)$$

Now, considering the partial derivative terms individually,

$$\frac{\partial}{\partial \vec{r}} \frac{1}{r^3} = -3r^{-4} \frac{\partial r}{\partial \vec{r}} \qquad (25)$$

$$\frac{\partial r}{\partial \vec{r}} = \frac{\partial \sqrt{\vec{r} \cdot \vec{r}}}{\partial \vec{r}} = \hat{r} \qquad (26)$$

The other partial term in Equation (22) is calculated,

$$\frac{\partial}{\partial \vec{r}} (\hat{r} \cdot \hat{p})^2 = 2(\hat{r} \cdot \hat{p}) \frac{1}{r} \frac{\partial}{\partial \vec{r}} (\hat{r} \cdot \hat{p}) = \frac{2}{r} (\hat{r} \cdot \hat{p})\hat{p} \qquad (27)$$

Thus, the disturbing acceleration from oblateness is

$$\vec{a}_{d_{20}} = \frac{-3\mu}{2r^4} C_{20}([1 - 3(\hat{r} \cdot \hat{p})^2]\hat{r} - 2[\hat{r} \cdot \hat{p}]\hat{p}) \qquad (28)$$

Now, for the computation of the ellipticity disturbing acceleration,

$$\vec{a}_{d_{22}} = 3\mu C_{22}[(\hat{r} \cdot \hat{s})^2 - (\hat{r} \cdot \hat{q})^2] \frac{\partial}{\partial \vec{r}} \frac{1}{r^3} + \frac{3\mu}{r^3} C_{22} \frac{\partial}{\partial \vec{r}} [(\hat{r} \cdot \hat{s})^2 - (\hat{r} \cdot \hat{q})^2] \qquad (29)$$

Noting results from Equations (22), (23), and (24),

$$\vec{a}_{d_{22}} = \frac{3\mu}{r^4} C_{22}(3[(\hat{r} \cdot \hat{s})^2 - (\hat{r} \cdot \hat{q})^2]\hat{r} + 2(\hat{r} \cdot \hat{s})\hat{s} - 2(\hat{r} \cdot \hat{q})\hat{q}) \qquad (30)$$

Combining the three disturbing accelerations from Equations (28) and (30),

$$\vec{a}_{d_m} = \frac{3\mu}{r^4}[C_{20}([1 - 3(\hat{r} \cdot \hat{p})^2]\hat{r} - 2[\hat{r} \cdot \hat{p}]\hat{p}) + C_{22}(3[(\hat{r} \cdot \hat{s})^2 - (\hat{r} \cdot \hat{q})^2]\hat{r} + 2(\hat{r} \cdot \hat{s})\hat{s} - 2(\hat{r} \cdot \hat{q})\hat{q})] \qquad (31)$$

The SRP and mass distribution disturbing accelerations are used in Cowell's formulation to describe the equations of motion. These are numerically integrated to simulate a short period motion about an asteroid. Motion about asteroid 101955 Bennu is considered. Table 1 represents required asteroid and satellite parameters [25]

Table 1: Asteroid and Spacecraft Parameters

| Parameter | Value |
|---|---|
| Oblateness $C_{20}$ | -0.0175 km$^2$ |
| Ellipticity $C_{22}$ | 0.0058 km$^2$ |
| Gravitational Parameter $\mu$ | $5*10^{-4}$ km$^3$/s$^2$ |
| Mass to Area Ratio m/A | 20 kg/m$^2$ |
| Albedo | 0.05 |
| Solar Radiation $P_\varphi$ | $1*10^8$ km$^2$/s$^2$m$^2$ |
| Asteroid vector | [1.12 0 0] AU |

The asteroid's inertia axes are chosen to be parallel to the x, y, and z axes of the asteroid's orbit for simplification (asteroid rotation is not taken into account due to the large rotational period relative to a hop). During simulations, the force is taken to be acting over the cross-sectional area of the robot evenly. Then, the net torque applied is calculated and included as the disturbing torque.

**Disturbances: Inflatable Deformation**

A non-linear finite element model was set up using Tsai-Belytschko membrane elements to capture the membrane's deformation behavior. A worst analysis was conducted to simulate the membrane colliding elastically with a rigid wall at an incoming velocity of 0.4 m/s. A constant volume inflation was used. The assumption this model makes is that the membrane is fully inflated at the time of collision.

Figure 10 shows the worst case expected deformation and stress plots of the membrane when it lands directly on the inflatable's side. A can be observed in this case a total deformation of about 60mm if expected to take place in the longitudinal direction. The corresponding worst-case change in inertia is accounted for in inertia uncertainties for numerical simulation.

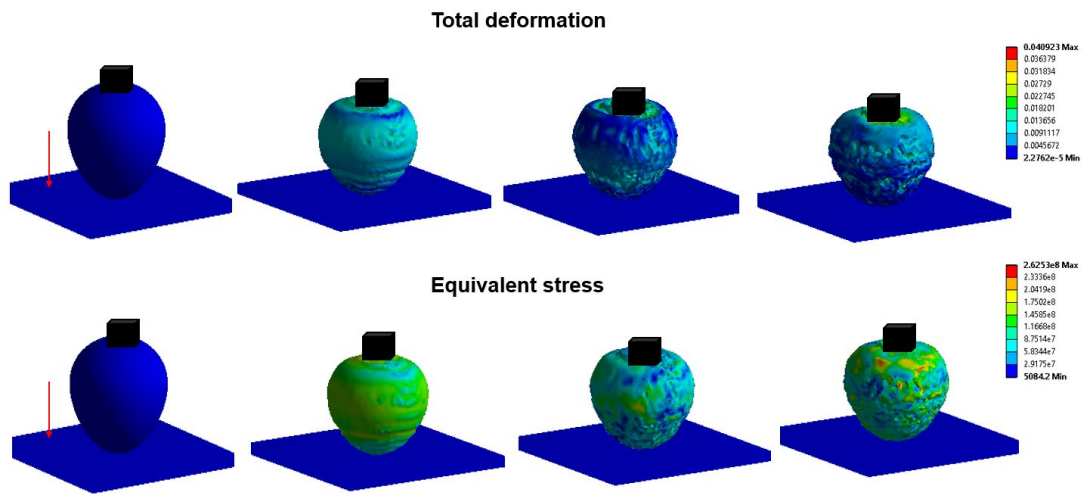

**Figure 10: Deformation and Stress Plots**

From the equivalent stress plots, it can be observed that stress concentrations begin to cause local bucking at the interface between the 1U CubeSat and the membrane causing to CubeSat to move into it. Table 2 describes analysis settings used:

**Table 2: Simulation Properties**

| Property | Value |
| --- | --- |
| Membrane Young's modulus (Mylar) | 760000 psi |
| Membrane Poisson's ratio (Mylar) | 0.38 |
| Membrane thickness | $1.27 \times 10^{-5}$ m |
| Internal membrane pressure | $2 \times 10^{-5}$ psi (similar to that expected by sublimates) |
| Initial membrane velocity | 0.4 m/s |

**Numerical Simulations**

To begin simulation, the initial state of the system is defined for a variety of cases. A 4$^{th}$ Order Runge-Kutta method is used to account for the non-linearity of quaternion rotation; they cannot be simply added together. MATLAB is used for simulations. Outputs of the simulation are plots of the quaternion and angular velocity evolution in time (Figure 11 and 12).

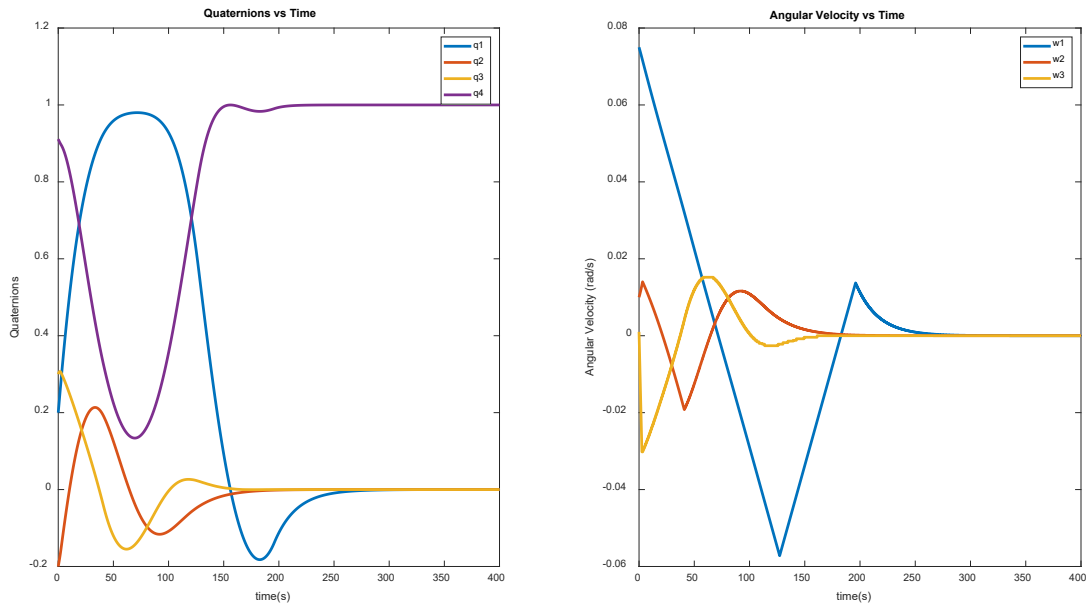

**Figure 11:** $q(0) = [0.2 \; -0.2 \; 0.3 \; 0.911]^T$, $\omega(0) = [0.075 \; 0.01 \; 0.001]^T rad/s.$

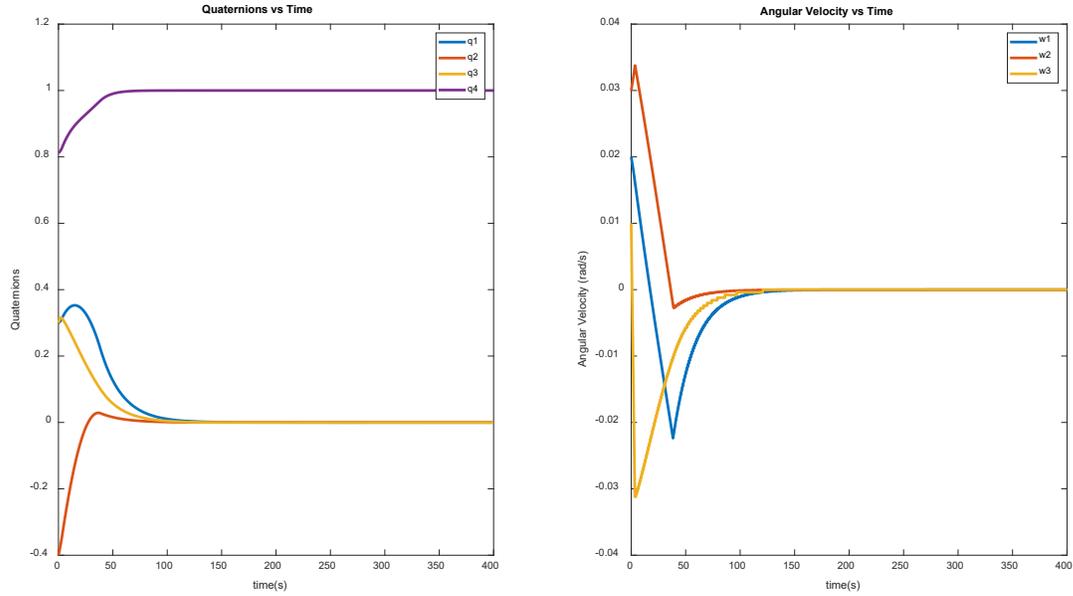

**Figure 12:** $q(0) = [0.3 \ -0.4 \ 0.3 \ 0.812]^T$, $\omega(0) = [0.02 \ 0.02 \ 0.01]^T rad/s.$

## UP-RIGHTING MANEUVERS

Should the robot not land upright, it will tilt over onto its side and rest on the surface with contact points on the inflatable surface and on the corner of the main structural cube. When this happens, the thruster chip is used to induce sliding motion against the regolith and liftoff of the robot. Assumptions made are that the regolith is homogeneous and flat (not rubble piles) and the contacts are points. A simplification to the model is planar motion, illustrated in Figure 13.

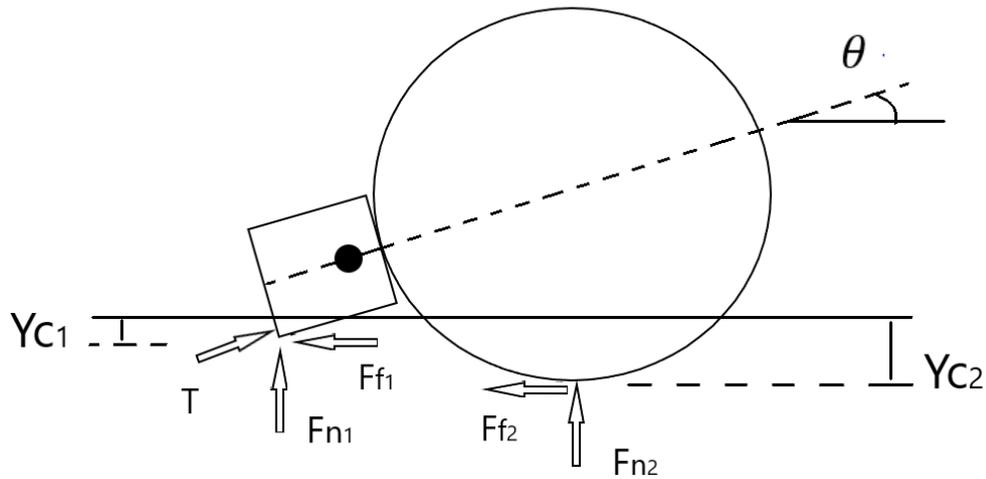

**Figure 13: AMIGO with Reaction Forces**

Drawing on knowledge from developments in the Hedgehog program [26], the system is modelled through Lagrangian mechanics [27]. The state of the planar robot is defined through generalized coordinates $g = (x, y, \theta)$. The motion of the robot is described by (32)

$$\frac{d}{dt}\frac{\partial T}{\partial \dot{g}} - \frac{\partial T}{\partial g} + \frac{\partial D}{\partial \dot{g}} + \frac{\partial V}{\partial g} = \frac{\partial \delta L}{\partial \delta g} \qquad (32)$$

Where T is the kinetic energy, D is dissipative energy, V is potential energy, and δL is the Lagrangian portion of the virtual work. These quantities can be defined through state $g$ and the two contact points, $(x_{c_1}, y_{c_1})$ for the contact of the corner and $(x_{c_2}, y_{c_2})$ for the inflatable contact point, both treated as points and assuming small impact depth:

$$T = \frac{1}{2}m(\dot{x}^2 + \dot{y}^2) + \frac{1}{2}J_x\dot{\theta}^2 \qquad (33)$$

$$V = mg_{ast}y + V_c \qquad (34)$$

Where $m$ is the mass of the robot, $J_x$ is the polar moment of inertia, $g_{ast}$ is local surface gravity, and $V_c$ is the contact potential energy,

$$V_c = \frac{1}{2}K(y_{c_1}^2 + y_{c_2}^2) \qquad (35)$$

Where K is the stiffness of the regolith. This term only is used when the point is actually in contact with regolith ($y_{ci} < 0$). The dissipative term is defined as

$$D = \frac{1}{2}C(\dot{y}_{c_1}^2 + \dot{y}_{c_2}^2) \qquad (36)$$

Where C is the damping coefficient of the regolith, and only considers $\dot{y}_{c_i} < 0$ (penetrating into the regolith). The virtual work is

$$L = u\theta + L_c \qquad (37)$$

Where $u$ is the control torque from the thrusters firing and $L_c$ is the work against friction,

$$L_c = F_{f_1}x_{c_1} + F_{f_2}x_{c_2} \qquad (38)$$

With slipping friction

$$F_{f_i} = -sign(\dot{x}_{c_i})\mu_d F_{n_i} \qquad (39)$$

Where $\mu_d$ is the dynamic friction coefficient and $F_{n_i}$ is the normal force at the contact point. The contact points are geometrically related to the center of mass and rotation angle $\theta$ when treated as a rigid body,

$$x_{c_1} = x - r_{c_1}\cos(\theta) \qquad (40)$$

$$y_{c_1} = y - r_{c_1}\sin(\theta) \qquad (41)$$

$$x_{c_2} = x + r_{c_2}\cos(\theta) \qquad (42)$$

$$y_{c_2} = y - r_{c_2}\sin(\theta) \qquad (43)$$

Where $r_{c_1}$ is the distance from the center of mass to the corner of the main structure, and $r_{c_2}$ is the distance from the center of mass to the contact point on the inflatable. From Equation (31), the contact dynamics can be described

$$m\ddot{x} = F_{f_1} + F_{f_2} \qquad (44)$$

$$m\ddot{y} + mg_{ast} + K[(y - r_{c_1}\sin(\theta)) + (y - r_{c_2}\sin(\theta))] + C[(\dot{y} - \dot{\theta}r_{c_1}\cos(\theta)) + (\dot{y} - \dot{\theta}r_{c_2}\cos(\theta))] = 0 \qquad (45)$$

$$J_x\ddot{\theta} + r_{c_1}\sin(\theta)\left[K(y - r_{c_1}\sin(\theta)) + C(\dot{y} - \dot{\theta}r_{c_1}\cos(\theta))\right] +$$
$$r_{c_2}\sin(\theta)\left[K(y - r_{c_2}\sin(\theta)) + C(\dot{y} - \dot{\theta}r_{c_2}\cos(\theta))\right] = u \qquad (46)$$

The goal of the up-righting control is to drive the robot upright ($\theta = \frac{\pi}{2}$) with the bottom of the robot resting on the surface (y = the distance from the center of gravity to the bottom of the robot). The adaptive sliding mode control law Equation (9) is modified,

$$S = \left[x, y_{c_1}, \theta - \frac{\pi}{2}\right]^T \qquad (47)$$

$$\xi = [\dot{x}\ \dot{y}\ \dot{\theta}] \qquad (48)$$

Euler's equation of motion for the planar case is used.

$$J_x = -J_x\omega^2 + u + L_{cont} \qquad (49)$$

Where $L_{cont}$ are the torques from contact forces.

## CONCLUSIONS

In this paper we presented an overview of the asteroid surface hopper AMIGO control system. Three mobility modes are controlled: hopping, attitude control during a hop, and up-righting maneuvers. The robot is able to hop to various locations on 101955 Bennu sized asteroids by thrusting an array of discretized nozzles on a MEMS chip. This thruster chip provides attitude control during a hop to ensure safe landing, limiting potential damage to the spacecraft and its inflatable structure. It is shown that the regulating adaptive sliding mode controller easily overcomes the disturbance torques from solar radiation pressure, asteroid oblateness and ellipticity. The controller is not highly dependent on the inertia of the robot, so deformations to the inflatable structure are well tolerated. As seen from Figures 9 and 10, the thrusters are sufficient to stabilize the robot at the identity attitude ($q4 = 1$) with no angular velocity in the presence of these disturbance torques and does so before the robot is able to land. In addition, the thrusters are able to correct positioning and avoid tip-over during landing.

Future work full focus on optimizing the fuel usage. The current system has some overshoot which requires extra propellant to counteract. This is more of an issue for larger errors in initial attitude from the identity attitude. This could be done through more rigorous optimization of the control parameters *β, p,* and *k₁*. In testing the control system, a buoyancy chamber will be designed and constructed to simulate milligravity environments. To further quantify regolith characteristics, an on-orbit centrifuge will simulate dust dynamics and micro-mobility platforms [28-30].

## REFERENCES

1. "Vision and Voyages for Planetary Science in the Decade 2013-2022." NASA, NASA, 18 Jan. 2018, solarsystem.nasa.gov/science-goals/about/.

2. Reddy, V., et al. "Mineralogy and Surface Composition of Asteroids." Asteroids IV, 2015, doi:10.2458/azu_uapress_9780816532131-ch003.

3. C. M. Hartzell, D. J. Scheeres, "Dynamics of levitating dust particles near asteroids and the Moon." J. Geophys. Res. Planets, 2013, 118, 116-126, doi:10.1029/2012JE004162.

4. R.M. Jones. The MUSES-CN rover and asteroid exploration mission. In 22nd International Symposium on Space Technology and Science, pages 2403-2410, 2000.

5. R. Allen, M. Pavone, C. McQuinn, I. A. D. Nesnas, J. C. Castillo-Rogez, Tam-Nquyen, J. A. Hoffman, "Internally-Actuated Rovers for All-Access Surface Mobility: Theory and Experimentation," IEEE Conference on Robotics and Automation (ICRA), St. Paul, Minnesota, 2012.

6. B. Hockman, A. Frick, I. A. D. Nesnas, M. Pavone, "Design, Control, and Experimentation of Internally-Actuated Rovers for the Exploration of Low-Gravity Planetary Bodies," Conference on Field and Service Robotics, 2015.

7. E. Dupius, S. Montminy, P. Allard, "Hopping robot for planetary exploration" 8th iSAIRAS, September 2005.

8. H. Khalita, S. Schwarz, E. Asphaug, J. Thangavelauthum, "Mobility and Science Operations on an Asteroid Using a Hopping Small Spacecrafts on Stilts" 42$^{nd}$ AAS GNC Conference, February, 2019, Breckenridge, CO.

9. J. Thangavelautham, M. S. Robinson, A. Taits, T. J. McKinney, S. Amidan, A. Polak, "Flying, hopping Pit-Bots for cave and lava tube exploration on the Moon and Mars" 2nd International Workshop on Instrumentation for Planetary Missions, NASA Goddard, Greenbelt, Maryland, 2014.

10. H. Kalita, R. T. Nallapu, A. Warren, J. Thangavelautham, "GNC of the SphereX Robot for Extreme Environment Exploration on Mars," Advances in the Astronautical Science, February 2017.

11. H. Kalita, R. T. Nallapu, A. Warren, J. Thangavelautham, "Guidance, Navigation and Control of Multirobot Systems in Cooperative Cliff Climbing," Advances in the Astronautical Science, February 2017.

12. L. Raura, A. Warren, J. Thangavelautham, "Spherical Planetary Robot for Rugged Terrain Traversal," IEEE Aerospace Conference, 2017.

13. Wilson, Jim. "What Is NASA's Asteroid Redirect Mission?" NASA, NASA, 16 Apr. 2015.

14. G. Wilburn, E. Asphaug, J. Thangavelautham, "A Milli-Newton Propulsion System for the Asteroid Mobile Imager and Geologic Observer (AMIGO)", IEEE Aerospace Conference, Big Sky, MT, 2019.

15. Wu, S.; Mu, Z.; Chen, W.; Rodrigues, P.; Mendes, R.; Alminde, L. TW-1: A CubeSat Constellation for Space Networking Experiments. In Proceedings of the 6th European CubeSat Symposium, Estavayer-le-Lac, Switzerland, 16 October 2014.

16. Bonin, G et al. CanX–4 and CanX–5 Precision Formation Flight: Mission Accomplished! In Proceedings of the 29th Annual AIAA/USA Conference on Small Satellites, Logan, UT, USA, 8–13 August 2015.

17. Underwood, C.I.; Richardson, G.; Savignol, J. In-orbit results from the SNAP-1 nanosatellite and its future potential. Philos. Trans. R. Soc. Lond. A Math. Phys. Eng. Sci. 2003, 361, 199–203.

18. Hejmanowski, N.J.C.A.; Woodruff, R.B. CubeSat High Impulse Propulsion System (CHIPS). In Proceedings of the 62nd JANNAF Propulsion Meeting (7th Spacecraft Propulsion), Nashville, TN, USA, 1–5 June 2015.

19. Robin, M.; Brogan, T.; Cardiff, E. An Ammonia Microresistojet (MRJ) for micro Satellites. In Proceedings of the 44th AIAA/ASME/SAE/ASEE Joint Propulsion Conference & Exhibit, Hartford, CT, USA, 21–23 July 2008.

20. Guo, Jian, et al. "In-Orbit Results of Delfi-n3Xt: Lessons Learned and Move Forward." Acta Astronautica, vol. 121, 2016, pp. 39–50., doi:10.1016/j.actaastro.2015.12.003.

21. H. Kalita, J. Thangavelautham, "Motion Planning on an Asteroid Surface with Irregular Gravity Fields," Advances in the Astronautical Sciences, February 2019.

22. W. Gao and J. C. Hung, "Variable structure control of nonlinear systems: A new approach," IEEE Trans. Ind. Electron., vol. 40, no. 1, pp. 45–55, Feb. 1993.

23. Zhu, Zheng, et al. "Adaptive Sliding Mode Control for Attitude Stabilization With Actuator Saturation." IEEE Transactions on Industrial Electronics, vol. 58, no. 10, 2011, pp. 4898–4907., doi:10.1109/tie.2011.2107719.

24. D.j., Scheeres. "Orbit Mechanics About Asteroids and Comets." Journal of Guidance, Control, and Dynamics, vol. 35, no. 3, 2012, pp. 987–997., doi:10.2514/1.57247.

25. Scheeres, D.j., et al. "The Geophysical Environment of Bennu." Icarus, vol. 276, 2016, pp. 116–140., doi:10.1016/j.icarus.2016.04.013.

26. R. G. Reid, L. Roveda, I. Nesnas and M. Pavone. (2014). "Contact Dynamics of Internally-Actuated Platforms for the Exploration of Small Solar System Bodies".

27. Lanczos, Cornelius. *The Variational Principles of Mechanics*. Dover Publications, 2005.

28. Asphaug, E., Thangavelautham, J., "Asteroid Regolith Mechanics and Primary Accretion Experiments in a Cubesat," 45th Lunar and Planetary Science Conference, 2014.

29. Lightholder, J., Thoesen, A., Adamson, E., Jakubowski, J., Nallapu, R., Smallwood, S., Raura, L., Klesh, A., Asphaug, E., Thangavelautham, J., "Asteroid Origins Satellite 1: An On-orbit CubeSat Centrifuge Science Laboratory," Acta Astronautica, Vol 133, 2017, pp. 81-94.

30. Thangavelautham, J., Asphaug, E., Schwartz, S., "An On-Orbit CubeSat Centrifuge for Asteroid Science and Exploration," IEEE Aerospace Conference 2019.

31. Y. Xu, K. W. Au, G. C. Nandy, H. B. Brown, "Analysis of Actuation and Dynamic Balancing for a Single Wheel Robot" *IEEE/RSJ International Conference on Intelligent Robots and Systems*, October 1998.

32. S. Dubowsky, K. Iagnemma, S. Liberatore, D. M. Lambeth, J. S. Plante, P. J. Boston, "A concept Mission: Microbots for Large-Scale Planetary Surface and Subsurface Exploration" *Space Technology and Applications International Forum*, 2005.